\documentclass[twocolumn,showpacs,preprintnumbers,amsmath,amssymb]{revtex4-1}

\usepackage{graphicx}
\usepackage{dcolumn}
\usepackage{bm}
\usepackage{float}

\begin{document}


\title{$Ab$ $initio$ electronic stopping power and threshold effect of channeled slow light ions in HfO$_{2}$}



\author{Chang-Kai Li$^{1,2}$, Feng Wang$^{3}$, Bin Liao$^{1,2}$, Xiao-Ping OuYang$^{5}$ and {Feng-Shou Zhang}$^{1,2,4,}$}
\email[]{Corresponding author. fszhang@bnu.edu.cn}
\affiliation{%
$^{1}$The Key Laboratory of Beam Technology and Material Modification of Ministry of Education, College of Nuclear Science and Technology, Beijing Normal University
Beijing 100875, China \\
$^{2}$Beijing Radiation Center, Beijing 100875, China\\
$^{3}$School of Physics, Beijing Institute of Technology, Beijing 100081, China\\
$^{4}$Center of Theoretical Nuclear Physics, National
Laboratory of Heavy Ion Accelerator of Lanzhou, Lanzhou 730000,
China\\
$^{5}$Northwest Institute of Nuclear Technology, Xi¡¯'an 710024, China\\
}%


\date{\today}

\begin{abstract}
We present $ab$ $initio$ study of the electronic stopping power of protons and helium ions
in an insulating material, HfO$_{2}$. The calculations are carried out in channeling conditions with different impact
parameters by employing Ehrenfest dynamics and real--time, time--dependent density
functional theory. The satisfactory comparison with available experiments demonstrates that this approach provides an accurate description
of electronic stopping power. The velocity-proportional stopping power is predicted for protons and helium ions in the low energy region, which conforms the linear response theory. Due to the existence of wide band gap, a threshold effect in extremely low velocity regime below excitation is expected. For protons, the threshold velocity is observable, while it does not appear in helium ions case. This indicates the existence of extra energy loss
channels beyond the electron--hole pair excitation when helium ions are moving through the crystal. To analyze it, we checked the charge state of
the moving projectiles and an explicit charge exchange behavior between the ions and host atoms is found. The missing threshold effect for helium ions is attributed to the charge transfer, which also contributes to energy loss of the ion.

\end{abstract}


\maketitle
\section{ INTRODUCTION}
Interaction of intruding ion with matter is of continuing interest, the deceleration force involved is mainly characterized by stopping power (SP) \cite{PhysRevB.16.115} or energy deposition per unit distance traveled through the material. This quantity can generally be decomposed into two parts. The
first one is the electronic stopping power \emph{$S_{e}$}, which arises from the excitation of the electrons of the target. The second one is the nuclear stopping
power \emph{$S_{n}$}
due to elastic Coulomb collisions with the nuclei of the target. For particles with velocities below the Fermi velocity of
the target, nuclear and electronic stopping are both relevant, and the result of the interaction is a collision cascade \cite{Averback1997Displacement}. Shortly before the particle stops eventually, a global stopping maximum denoted Bragg peak occurs. Thus, studying the energy transferred from slow ions to the target material is at the heart of modern technologies, such as nuclear fission/fusion reactors \cite{Odette2005,PhysRevB.93.245106}, semiconductor devices for space missions \cite{ZELLER19952041} and cancer therapy based on ion beam radiation \cite{Obolensky2008Ion,PhysRevB.94.041108}.

The interest in modeling the SP
of ions with velocities between 0.1 and 1 atomic units (a.u.
hereafter) has fueled a huge amount of research \cite{a}. In this regime the electronic component is generally dominant and the
electronic energy loss is predominantly due to interaction
with valence electrons \cite{PhysRevLett.85.2825}. The electronic stopping power is a crucial quantity for ion irradiation: it governs the deposited heat, the damage
profile, and the implantation depth. Ever since the discovery of \emph{$S_{e}$},
various models and theories have been proposed to calculate $S_{e}$ depending on the energy regime of the ion. For swift ions, based on the assumption that the atoms of the target are classical oscillator, Bohr \cite{bohr,bohr1} suggested a formula for the $S_{e}$; employing
the first Born approximation, Bethe \cite{bethe} has introduced the
first calculations of inelastic and ionization cross section; the
Bloch correction \cite{bloch} provides a convenient link between
the Bohr and the Bethe scheme. On the other hand, for slow particles, Fermi and Teller \cite{Fermi1947The} using
electron gas models had reported $S_{e}$ for
various targets; a more general treatment of the $S_{e}$ applicable to the whole velocity regime was later developed by Lindhard \cite{Lindhard1954ON} through linear dielectric theory.

Energetic ions penetrate great depth along channels between
low--index crystallographic planes, moderating through collisions with electrons,
until finally they hit a host atom initiating a cascade of atomic displacements. This $channeling$ phenomenon
has been exploited in many important applications such as ion implantation and depth profiling. Glancing collisions with host atoms confine the trajectory of a channelling ion, so most of
its energy is lost through electronic excitation. Based on the free electron gas (FEG) model, \emph{$S_{e}$} is predicted to be \emph{$S_{e}$} $\propto$ $v$ for $v$ $<$ 1 atomic units (a.u. henceforth) \cite{Fermi1947The}. This simple relation has been verified experimentally in many $sp$--bonded metals \cite{Race2010The,Vald1993Electronic,Mart1996Energy,Pitarke1999Band}.

A different behavior is expected in materials that have a finite minimum excitation energy T$_{min}$, such as noble gas and wide band--gap insulators, given the finite energy required to excite outmost electrons. This finite minimum excitation energy is expected to suppress the energy dispassion due to electron--hole pair excitation at ion energies of several keV \cite{Fermi1947The}, which results in a threshold effect of SP with respect to the ion velocity \cite{PhysRevLett.103.113201}. Instead, however, no threshold effect was originally observed in most systems \cite{PhysRevLett.79.4112,PhysRevA.64.012902,PhysRevLett.93.042502}, with the exception that a threshold velocity of $v \simeq 0.2$ a.u. in LiF under grazing incidence \cite{PhysRevLett.81.4831}. Recent theoretical and experimental research managed to measure velocities in extremely low velocity regime ($v$ $\leq$ 0.1 a.u) displaying a clear velocity threshold \cite{PhysRevA.76.040901,PhysRevA.81.022902,PhysRevLett.103.113201,PhysRevB.91.125203,PhysRevA.89.022707}.

In this work, we propose to investigate the \emph{$S_{e}$} behavior of slow ions shooting through a large band--gap insulator HfO$_{2}$ ($E_\mathrm{g}$ $\approx$ 5.5 eV)
and also its threshold effect under channeling condition. For this purpose, we perform non--adiabatic dynamic simulations to mimic the intruding ion as it is propagating
through the system by means of time--dependent density--functional theory (TDDFT).
The explicit time evolution of the electronic
states of the host crystal and the kinetic energy loss of the moving ions are monitored. The key
quantity of interest \emph{$S_{e}$} can be extracted from the change of kinetic projectile energy using the thickness of the target.

This article is outlined as follows. In Sec. II, we briefly introduce the
theoretical framework and the computational details. Results are presented
and discussed in Sec. III, where we concentrate on the analysis of \emph{$S_{e}$} and its threshold effect. In the
end, conclusions are drawn in Sec. IV. Henceforth, if no special reservation is made, Hartree atomic units with $m = |e| = \hbar = 1$ are used in this paper.

\section{ MODEL AND METHODS}

The collision behaviors of intruding ions with the host nuclei and electrons are characterized  by the Ehrenfest coupled electron--ion dynamics combined with time--dependent density--functional theory (ED--TDDFT) \cite{PhysRevLett.108.225504,Horsfield2004Power,Gross1996,Calvayrac2000Nonlinear,Alonso2008Efficient,Page2008The}. In this model, electronic degrees of freedom are treated quantum--mechanically within the time--dependent Kohn--Sham (KS) formalism, while the ions are handled classically. This
method allows for an excited electronic state \emph{ab initio} molecular dynamics
(AIMD) simulation. The ED--TDDFT can, in general, be defined by the following coupled differential equations:

\begin{eqnarray}
M_{I}\frac{d^{2}\vec{R}_{I}(t)}{dt^{2}}=&&-\int\Psi^{*}(x,t)[\nabla_{I}\hat{H}_{e}(\vec{r},\vec{R}(t))]\Psi(x,t)dx \nonumber \\
 &&-\nabla_{I}\sum_{I \neq J}\frac{Z_{I}Z_{J}}{|\vec{R}_{I}(t)-\vec{R}_{J}(t)|},
\end{eqnarray}
\begin{equation}
i\frac{\partial\Psi(x,t)}{\partial t}=\hat{H}_{e}(\vec{r},\vec{R}(t))\Psi(x,t),
\end{equation}
here $M_{I}$ and $Z_{I}$ denote the mass and charge of the \emph{I}th nuclei, respectively, and $\vec{R}_{I}(t)$ describes the corresponding ionic position vector.
$\Psi(x,t)$ is the many--body electron wave function in the time domain, for which
we define $x\equiv \{x_{j}\}_{j=1}^{N}$, with $x_{j}\equiv(\vec{r}_{j},\sigma_{j})$,
where the coordinates $\vec{r}_{j}$ and the spin $\sigma_{j}$ of the \emph{j}th electron
are implicitly taken into account. Here $N$ is the number of electrons of the system.

The electronic Hamiltonian is expressed as $\hat{H}_{e}(\vec{r},\vec{R}(t))$, which
depends on the instantaneous distribution of the positions of all the nuclei, $\vec{R}(t)\equiv\{\vec{R}_{1}(t), \cdots, \vec{R}_{M}(t) \}$ ($M$ is the number of nuclei of the system), and of all the electrons $\vec{r}$; thus, it basically consists of the kinetic energy of electrons, the electron--electron potential and the electron--nuclei potential, which can be formulated as
\begin{eqnarray}
\hat{H}_{e}(\vec{r},\vec{R}(t))=&&-\sum_{j}^{N}\frac{1}{2}\nabla^{2}_{j}+\sum_{i<j}\frac{1}{|\vec{r}_{i}-\vec{r}_{j}|} \nonumber \\
&&-\sum_{iI}\frac{Z_{I}}{|\vec{R}_{I}-\vec{r}_{i}|},
\end{eqnarray}

To solve Eq.\,(1) for the motion of the nuclei, one has to obtain knowledge of $\Psi(x,t)$,
which typically causes the problem to be intractable. For this reason, we write the force that acts on each nucleus in terms of the electronic density $n(\vec{r},t)$, which is the
basic variable of ED-TDDFT (see Ref.~\cite{Castro2012130} for
a detailed description). As a consequence, Eq.\,(1) can be rewritten as
\begin{eqnarray}
M_{I}\frac{d^{2}\vec{R}_{I}(t)}{dt^{2}}=&&-\int n(\vec{r},t)[\nabla_{I}\hat{H}_{e}(\vec{r},\vec{R}(t))]d\vec{r} \nonumber \\
&&-\nabla_{I}\sum_{I \neq J}\frac{Z_{I}Z_{J}}{|\vec{R}_{I}(t)-\vec{R}_{J}(t)|},
\end{eqnarray}
where the instantaneous density $n(\vec{r},t)$ is given by the sum of all individual
electronic orbitals, i.e.,
\begin{equation}
n(\vec{r},t)=\sum_{i=1}^{occ}| \varphi_{i}(\vec{r},t)|^{2},
\end{equation}
with $\varphi_{i}(\vec{r},t)$ being occupied KS orbital for the $i$th electron.

Similarly, to obtain $n(\vec{r},t)$ explicitly, instead of solving Eq.\,(2), we make use
of the corresponding time--dependent KS equations, which provides an
approximation to $n(\vec{r},t)$,
\begin{eqnarray}
[-\frac{1}{2}\nabla^{2}&&-\sum_{I}\frac{Z_{I}}{|\vec{R}_{I}(t)-\vec{r}|}+\int\frac{n({\vec{r}^{'}},t)}{|{\vec{r}-\vec{r}^{'}}|}d\vec{r}^{'} \nonumber \\
&&+V_{xc}(\vec{r},t)]\varphi_{i}({\vec{r}},t)=i\frac{\partial\varphi_{i}({\vec{r}},t)}{\partial t},
\end{eqnarray}
where $V_{xc}(\vec{r},t)$ is the time--dependent exchange-correlation potential, for which we use the adiabatic local density approximation with Perdew--Wang analytic parametrization \cite{PhysRevB.45.13244}. The other three terms on the left hand side of Eq.\,(6) are, in order, the electronic kinetic, the electron--nucleus potential, and the Hartree potential. In this model, the potential
energy and forces acting upon the ions are calculated $on$
$the$ $fly$ as the simulation proceeds. It should be noted that, in ED--TDDFT, transitions between electronic adiabatic
states are included, and the coupling of
the
adiabatic states with the nuclei trajectories is also considered \cite{edtddft}. Thus, it allows
$ab$ $initio$ molecular dynamics simulation for excited electronic states and opens a way
to study the electron transfer between the ion and the target
electrons during the collision \cite{chargetransfer}.
In the present work, since we are mainly interested in the low velocity regime that is well below the core--electron excitation threshold, only valence electrons of host atoms are considered throughout this work.
The coupling of valence electrons to ionic cores is described by using
norm--conserving Troullier--Martins (TM) pseudopotentials \cite{PhysRevB.43.1993}.

In order to investigate electron excitation and electron distribution after the collision, the electron distribution in the conduction band is defined by the projection of the time--dependent wave function to the initial particle states in the conduction band \cite{PhysRevB.77.165104,Jiao2013823},

\begin{equation}
n_{occ}(m,k)=\sum_{n'}|\langle\varphi_{mk}|\varphi_{n'k}(T)\rangle|^{2},
\end{equation}
where $m$ represents the KS orbital index and $k$ is the Bloch wave number.

The projectile is initially placed outside the crystal and a TDDFT \cite{PhysRevB.54.4484,PhysRevB.93.035128} calculation was completed to obtain the converged ground state results that are required as the initial condition for subsequent evolution with the moving projectile.
Once the convergence of the ground state is achieved,
the projectile is then released with the given initial velocity;
meanwhile, the KS orbitals are propagated
through the time--dependent KS equations by employing the approximated enforced time reversal symmetry method \cite{Castro2004Propagators}.
The ionic motion is obtained via the numerical solution of Eq.\,(4)
by applying Verlet's algorithm.
Each simulation of the ion--solid collision consists of a well-defined direction of the projectile shooting the HfO$_{2}$ thin film with experimental density.
The calculations were carried out by using the OCTOPUS \emph{ab initio} code package \cite{marques2003octopus,Castro2006Octopus}. In the present work,
the external potential, electronic density, and KS orbitals are discretized in a
set of mesh grid points with a uniform spacing of 0.18 {\AA} along the
three spatial coordinates in the simulation box. To avoid artificial reflections of the electronic
wavefunctions from the boundary, we use a complex absorbing potential boundary \cite{Wang2011A} during the collision process.

Physical picture of this study is that: the projectiles are directed along negative \emph{z} axis with a given velocity. Ionic motion of target atoms is neglected by fixing the host ions in the equilibrium positions because it is supposed to play a marginal role over the total simulation time that is limited to several femtoseconds \cite{Correa2012Nonadiabatic}. \emph{$S_{e}$} was investigated under channeling condition, where the projectiles do not encounter the target nuclei directly. The nuclear contribution to the stopping power, therefore, is negligibly small and can even be completely suppressed when the host atoms are frozen in the equilibrium positions.

Supercell size is chosen so as to reduce the finite size effects while maintaining controllable computational demands. In present work, a FCC structured 2$\times$2$\times$2 conventional cell comprising 32 Hf and 64 O atoms is employed,
the lattice constants we choose is 5.11 {\AA}, which is identical to the measured value \cite{Wyckoff}. To ensure the stability of the computation, we use a time step of 0.001 fs. It should be noted that
the numerical parameters have been carefully examined, and the chosen parameters are found to be a compromise between the convergence of results and the efficiency of the calculation.

\section{ RESULTS AND DISCUSSION}
\emph{$S_{e}$} along the middle axes of three different
channels is investigated. We show in Fig.~1 the calculated results for \emph{$S_{e}$} of protons and helium ions in HfO$_{2}$ with velocity range 0.1 -- 1.0 a.u., together with the experimental data \cite{Primetzhofer2014100}. For the sake of
comparison, the predictions obtained from the SRIM--2013 data base
are also plotted. It
should be noted that SRIM results are obtained semi--empirically by
averaging over a number of different incident directions with distinct impact parameters; thus, it does not explicitly account for the
channeling conditions studied in our calculations. For this reason,
the calculated results are expected to follow a qualitative trend of
SRIM data, which has been shown in several previous theoretical studies \cite{Schleife2015Accurate,Correa2012Nonadiabatic,PhysRevB.94.155403}.

In Fig.~1(a) we find a pronounced agreement between the calculated values and experimental data for \emph{$S_{e}$} of protons along the middle axes of $<$100$>$ and $<$111$>$ channels, which shows the power of the TDDFT technique to accurately reproduce electronic stopping power in realistic system. However, in order to
obtain the electronic stopping power with high precision, an ensemble average over numerous projectile paths
needs to be taken until satisfactory convergence is reached, but this is beyond the scope of this work. As can be seen in Fig.~1, the \emph{$S_{e}$} behavior of protons and helium ions in HfO$_{2}$ are generally velocity--scaling, which conforms the expected energy dissipation mechanism caused by electron--hole pair excitation. An interesting phenomenon we find in Fig.\,1 is that the threshold effect for protons and helium shows a distinctly different trait. Considering the velocity-proportionality assumption, if we extrapolate the calculated data to zero, the threshold velocities for protons shown in Fig.~1(a) are 0.07 a.u., 0.10 a.u., and 0.04 a.u. in the $<$100$>$, $<$110$>$, and $<$111$>$ directions, respectively. For the \emph{$S_{e}$} behavior of helium ions shown in Fig.\,1(b), no threshold effect is found.
\begin{figure}[htp]
  \centering
  \includegraphics[width=10cm,height=8cm]{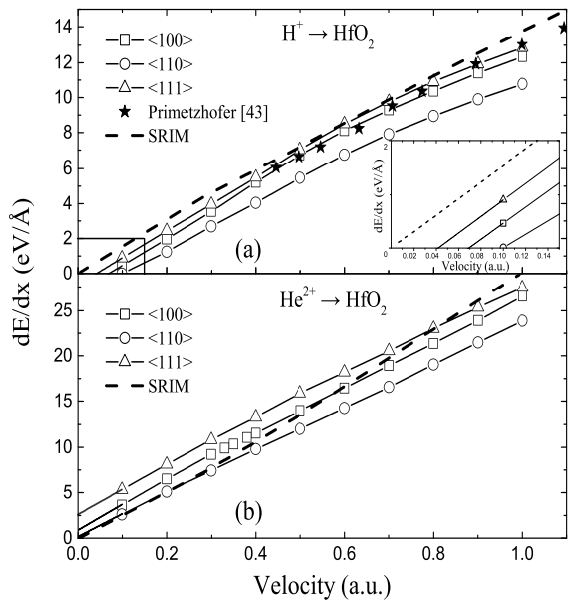}\\
  \caption{Electronic stopping power for protons (a) and helium ions (b) versus velocity along the middle axes of $<$100$>$, $<$110$>$ and $<$111$>$ channels of HfO$_{2}$. The lines are guides to the eye. The inset in (a) is enlargement of the main figure.}
\end{figure}

\subsection{Threshold effect in \emph{$S_{e}$} of HfO$_{2}$}
To investigate the different threshold effects between the protons and helium ions, we have explored the possible energy dissipation channels.
In the present work, we find \emph{$S_{e}$} is related
to charge transfer, which is an additional energy dissipation channel besides electron--hole pair excitation. Many
mechanisms may contribute to the charge transfer. Besides the direct transitions such as excitation, ionization,
and capture \cite{PhysRevA.84.052703}, the Auger process between the host atoms
and ions also plays a pronounced role, in which an electron jumps from the valence band of the host atom to an
ion bound state and vice versa. The energy released in
such transitions is balanced by an electronic excitation
in the medium or on the projectile \cite{DrezMuino20038}. Other possible
mechanisms are resonant charge transfer and radiative
decay processes of the projectiles. Since pseudopotentials are adopted in the present TDDFT simulation,
Auger processes following the inner--shell vacancy can not
be considered. Nevertheless, because the kinetic ion energy in the present work is restricted to 25 keV/u and
lower, according to the interpretation in Ref.~\cite{PhysRevA.87.032711}, direct transition mechanisms are dominant in such a low
velocity regime, Auger processes following the inner--shell
vacancy make a minor contribution to charge transfer.

Generally, both the neutralization and re-ionization of the ion included in the charge transfer process
contribute to the decreasing of ion's kinetic energy, due to the promotion of electronic states of either the host atoms or the projectile itself.
 Pe\'{n}alba $et$ $al.$ \cite{Pe1992Stopping} reported that charge transfer is an important energy loss channel, especially for projectile around stopping maximum. For protons with $v =$ 1 a.u. in Aluminum, charge transfer accounts for 15$\%$ of total SP. However, it has not been considered in the original SP theory that accounts for linear velocity dependence. The different threshold effect between proton and helium ion in Fig.\,1 is studied by checking their charge transfer behaviors in low velocity regime.

As a first step, the time evolution of a helium ion moving through the $<$100$>$ channel for a given velocity of 1.0 a.u. is visualized in Fig.\,2. Four snapshots covering the entire collision process are
presented. Before entering the crystal ($t$ = 0.059 fs), the helium ion projectile is a bare ion [Fig.\,2(a)]; when the helium ion is getting close to the crystal ($t$ = 0.095 fs) and penetrating along the channel ($t$ = 0.486 fs), it exchanges charge with the host atoms [Figs.\,2(b) and (c)]; after traversing the HfO$_{2}$ film ($t$ = 0.628 fs); the exiting ion still retains some captured electrons [Fig.\,2(d)]. The
example in Fig.\,2 could be qualitative evidence for the charge transfer during the collisions. The gray region is the change in electron distribution caused by the intruding ion.
\begin{figure}[htp]
  \centering
  \includegraphics[width=9cm,height=8cm]{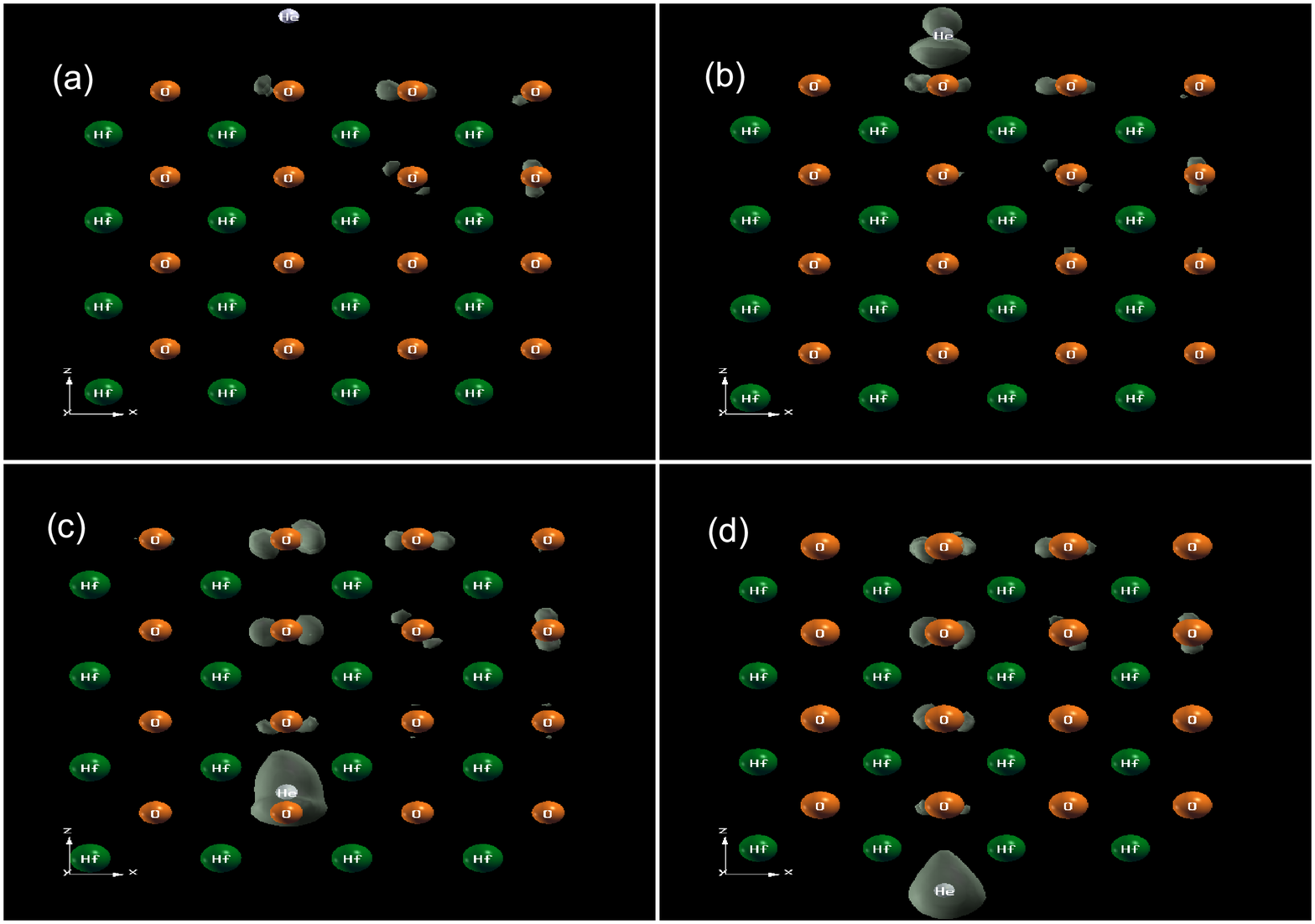}\\
  \caption{(Color online) Snapshots of time evolution of the electron density change caused by a He$^{2+}$ ion moving through HfO$_{2}$ crystal with $v$ = 1.0 a.u. along the middle axis of $<$100$>$ channel (side view). (a) $t = 0.059$ fs, the ion is above the crystal. (b) $t = 0.095$ fs, the ion is entering the channel. (c) $t = 0.486$ fs, the ion is penetrating along the channel. (d) $t = 0.628$ fs, the leaving ion retains some electrons after the collision. The
gray regions are the change of electron distribution caused by
the intruding ions.}
\end{figure}
\begin{figure}[htp]
  \centering
  \includegraphics[width=10cm,height=8cm]{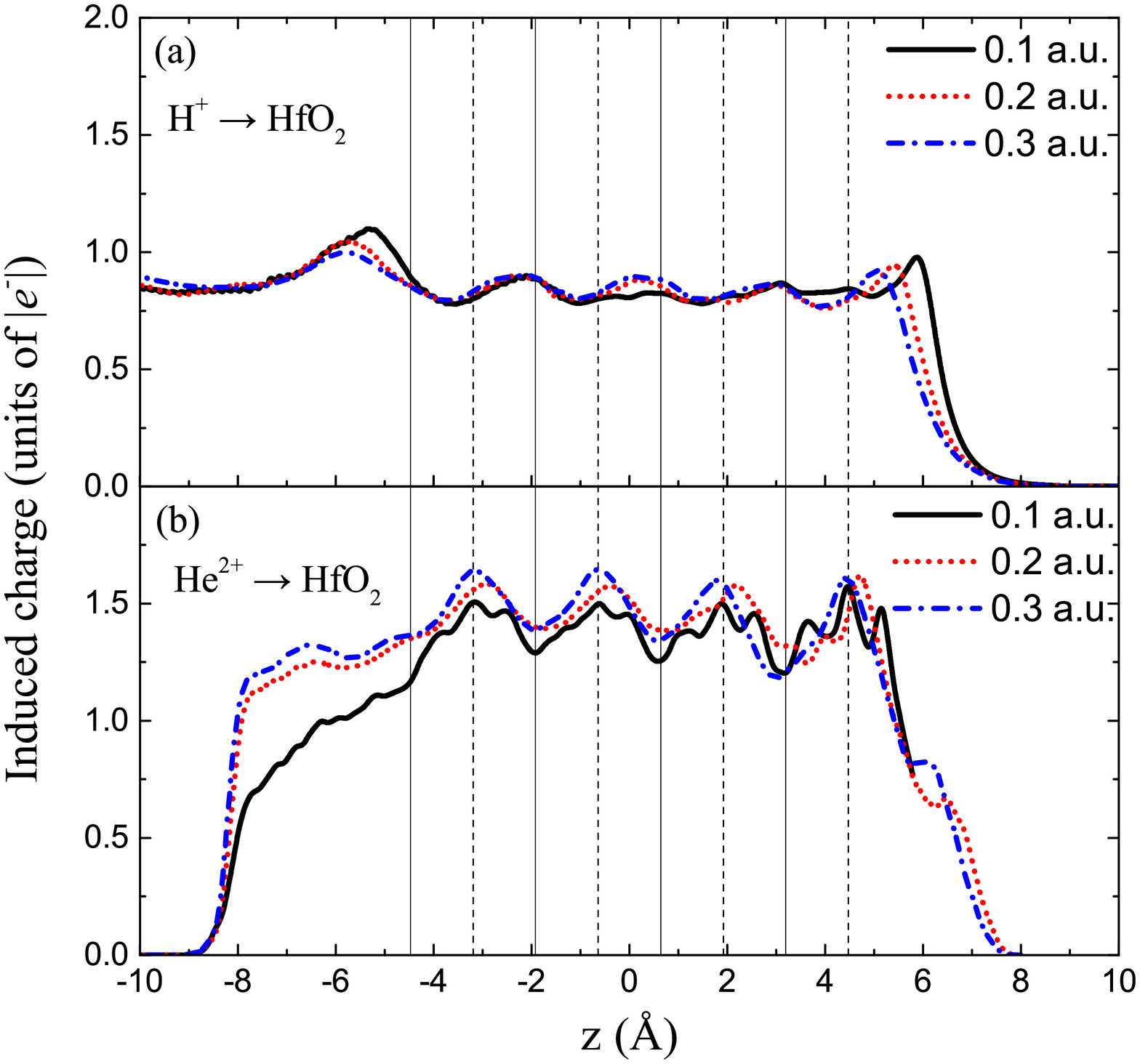}\\
  \caption{(Color online) Electrons captured by protons (a) and helium ions (b) versus \emph{z} coordinates in low velocity regime along the middle axis of $<$100$>$ channel. The vertical dashed lines are the positions of O layers, the vertical solid lines are the positions of Hf layers.}
\end{figure}

The electron density change induced by the helium ion in real time is quantified by integrating the valence charge density within the volume around the projectile ion with a radius of 1.26 {\AA} in the time--dependent calculation, from which the ground state electron density of the target in the corresponding volume has been subtracted and we thus obtained the number of induced electrons in real time. In the present work, we deem this quantity as the charge transferred to the projectiles. A point should be noted is that free electrons caused by electron scattering process may also be included in such approximation, where
electrons pile up close to projectile due to the attractive interaction between electrons and the ions \cite{ECHENIQUE1981779}. The choice of 1.26 {\AA} as the integration radius is a compromise
between various factors. In this study, we are
interested in learning the real--time electron occupying
the intruding ion orbitals, and we get it through the discrepancy of density, i.e., the change in electron distribution around the ion between the time--dependent and the
ground state calculations. In theory, a larger integration
radius can be more effective to fully take a variety of
mechanisms and also the highly occupied orbits into account. However, at the same time, it may include more free
electrons and excited state electrons of the host atoms
that do not belong to the ion and also more excited state
electrons caused by the former steps, as shown in Fig.~2.

Electrons induced by the H$^{+}$ and He$^{2+}$ ions with very low velocities moving along the middle axis of $<$100$>$ channel are presented in Fig.\,3; the periodic variation in induced electron reflects
the periodicity of the crystal. As can be seen, the charge exchange between the projectile and host atoms takes place alternately along their trajectories. Generally, protons show less active charge exchange behavior during the collision than helium ions. In the present work, we found charge transfer of protons is sensitive to ions velocity, it becomes less and less obvious as the velocity decreases in the extremely low velocity regime (not shown). While, charge transfer of helium ions shows a weak correlation with velocity. Considering charge transfer accounts for a noticeable share of \emph{$S_{e}$}, in the extremely low velocity regime the contribution to \emph{$S_{e}$} from electron--hole pair excitation becomes feeble and charge transfer may be the dominant energy loss channel. The different threshold effect shown in Fig.~1 can be attributed to the different charge transfer behaviors between two kinds of ions in extremely low velocity regime.

To depict the electron--hole pair excitation, we calculated the electron occupation number
distribution in the conduction band for protons at a given velocity of 0.01 a.u. along the middle axis of $<$110$>$ channel and 0.5 a.u. along the middle axis of $<$100$>$ channel. Such choices are made with the consideration to compare the cases with and without explicit electron--hole excitation, because 0.01 a.u. in $<$110$>$ channel is lower than the threshold velocity and 0.5 a.u. in $<$100$>$ channel is higher than threshold velocity. Results are shown in Fig.~4. The
excitation energy of the crystal has been extended to 7 eV
in the conduction band by adding empty KS obtials in the simulation. We can see clearly that the excited
electrons are distributed broadly in the considered energy
range for the proton at velocity of 0.5 a.u. in $<$100$>$ channel. However, the electrons occupation in conduction band is implicit for the
proton at velocity of 0.01 a.u. in $<$110$>$ channel, and the number of excited electrons in this energy
range is much smaller than that of the 0.5 a.u. proton.

As already alluded to in the previous paragraph, electrons excitation from valence band to the conduction band
are suppressed by the energy gap. It should be noted that charge transfer discussed above also contributes to electrons excitation. The implicit electrons excitation for 0.01 a.u. proton also demonstrates the missing of charge transfer in very low velocity regime. This means the projectile
ion does not lose any energy to the target electron subsystem
when the velocities are below a certain threshold, resulting
in a threshold effect in the \emph{$S_{e}$} dependence on the velocity. Although the empty KS level does not provide the
exact description of the real excited states of the nanostructure,
this can be considered as the first approximation to the true
excitations of the system \cite{PhysRevLett.90.043005,PhysRevLett.95.163006}.

\begin{figure}[htp]
  \centering
  \includegraphics[width=9cm,height=8cm]{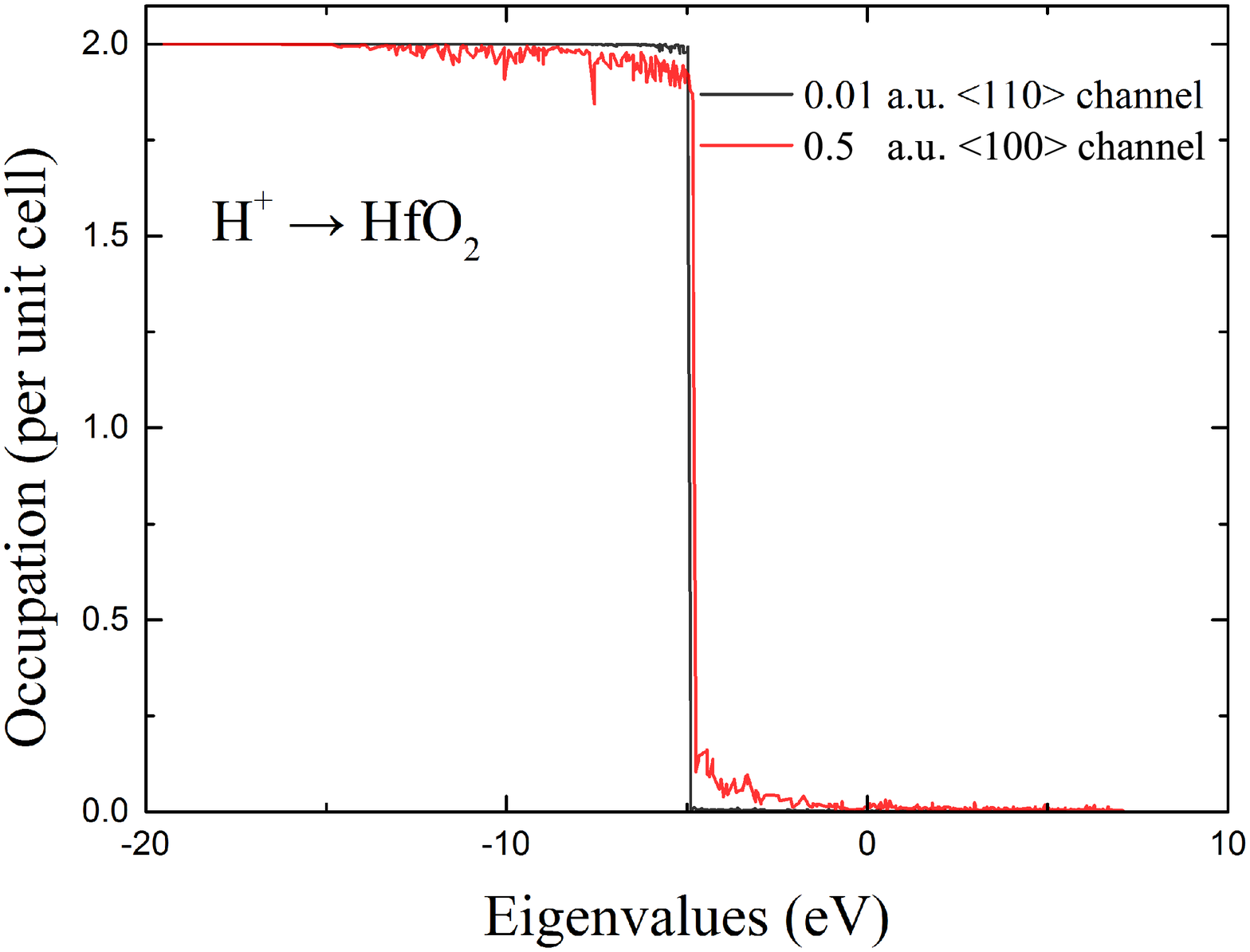}\\
  \caption{ Electron occupation number distribution after the collision ends. The left side of the vertical lines shows the occupation of the ground state and the right side shows the excited states. The black line and red one are cases of protons at 0.01 a.u. in $<$110$>$ channel and 0.5 a.u. in $<$100$>$ channel, respectively. See more details in the text.}
\end{figure}

\subsection{impact parameter dependence of charge transfer and \emph{$S_{e}$}}
Since the occupied He-1$s$ level is strongly affected by the interaction distance \cite{Wang2001Low,Wethekam2008Face,Monreal1989Channeling1,Primetzhofer2013Local}. So charge transfer behavior may differ for different trajectories. To investigate the effect of impact parameter on charge exchange and threshold velocity, we show in Figs.\,5 and 6 the \emph{$S_{e}$} behavior and the corresponding charge transfer profiles versus \emph{z} coordinates for a given velocity of 0.3 a.u. along two trajectories in $<$100$>$ channel and three trajectories in $<$110$>$ channel. The trajectories are chosen to sample different impact parameters within the channel. In the present work, the value of impact parameter is defined as the closest distance to any of Hf atoms along the ion trajectory. The trajectories along the middle axes of the channel have the highest impact parameters and trajectories close to Hf atoms have the lowest
impact parameters. For the five trajectories in Fig.~5, the impact parameters are in order 1.279 {\AA}, 0.904 {\AA}, 1.809 {\AA}, 1.357 {\AA} and 0.904 {\AA}.

As can be seen, no threshold effect can be found in $<$100$>$ channel in both the center and off--center channeling cases shown as trajectories 1 and 2, respectively. While, threshold effect is observed along the trajectories with low impact parameters in the $<$110$>$ channel, i.e. trajectories 4 and 5, the threshold velocity are 0.05 a.u. and 0.15 a.u., respectively. We notice that the charge transfer behavior shown in Fig.~6 is active for the trajectory in $<$100$>$ channel with low impact parameter. While, the amplitudes of charge transfer are much smaller along the trajectories in $<$110$>$ channel with low impact parameters, and trajectory 5 shows a even less evident charge transfer behavior than trajectory 4. This result consist with the threshold effect shown in Fig.~5, which indicates that an active charge transfer may cause vanish of the threshold.

\begin{figure}[htp]
  \centering
  \includegraphics[width=10cm,height=8cm]{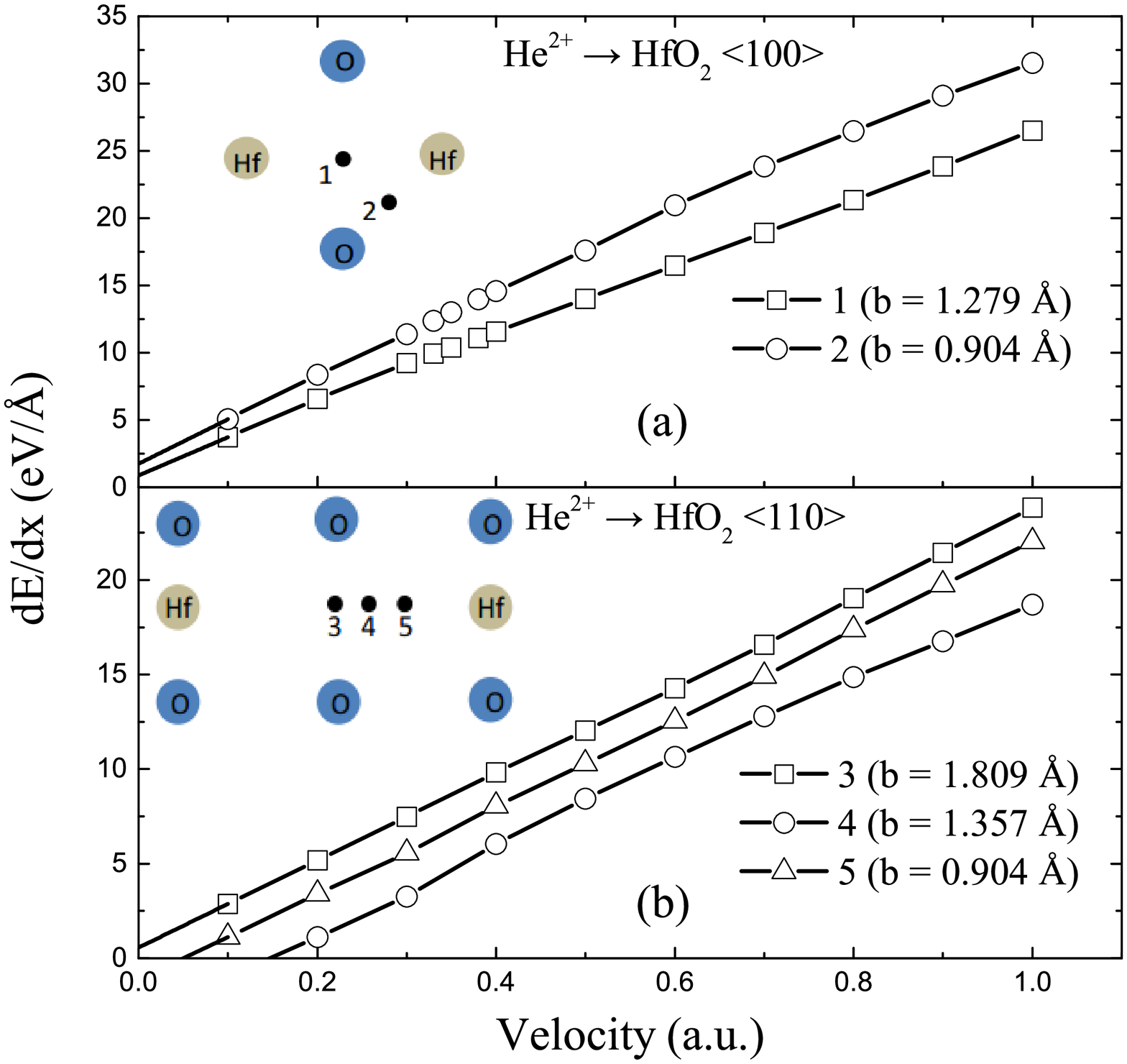}\\
 \caption{(Color online) Electronic stopping power of helium ions in HfO$_{2}$
as a function of the velocity along two trajectories in the $<$100$>$ channel (a) and three trajectories in the $<$110$>$ channel (b). The lines are guides to the eye. The inset shows
a sectional view of the $<$100$>$ channel and the trajectories. The gray
circles and the blue ones represent host atoms in different transverse planes (defining
the channel), while the black circles show the projectile positions for
different impact parameters. For the five trajectories impact parameters are in order 1.279 {\AA}, 0.904 {\AA}, 1.809 {\AA}, 1.357 {\AA} and 0.904 {\AA}.}

\end{figure}

\begin{figure}[htp]
  \centering
  \includegraphics[width=9cm,height=8cm]{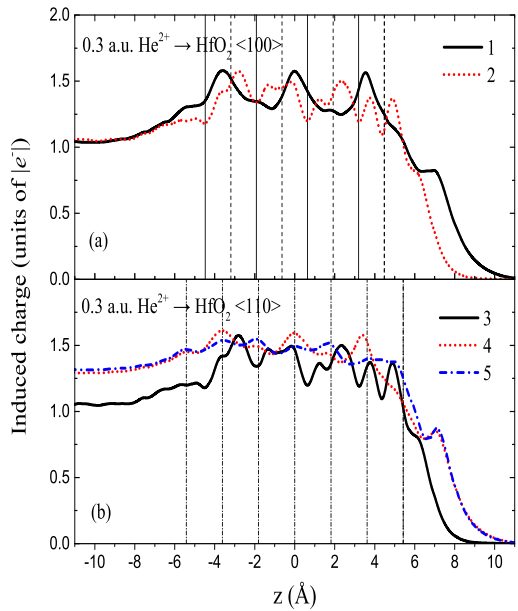}\\
  \caption{ Electron capture by He$^{2+}$ along trajectories with different impact parameter in the $<$100$>$ channel (a) and in the $<$110$>$ channel (b) for a given velocity of 0.3 a.u.. The curves are corresponding to trajectories in Fig.\,5. The vertical dashed lines and solid lines in (a) are the positions of O layers and Hf layers, respectively. The vertical dash-dotted lines in (b) are positions of crystalline layers comprising both O and Hf atoms.}

\end{figure}

The instantaneous energy loss rate of the projectile ion in a condensed matter system often depends strongly on the
specific path taken by the ion and its proximity to atoms
and bonds over the course of the trajectory. In the present work, we find \emph{$S_{e}$} is remarkably impact parameter--dependent, results along different trajectories
have different magnitudes.
To understand the dependence of the \emph{$S_{e}$} on the impact parameter,
we show in Fig.\,7 the ground state electronic density and the axial force along  along three different ion trajectories (shown as black circles in Fig.~5(b)) in $<$110$>$ channel for helium ions at a given velocity of 0.3 a.u.. The density values are obtained by averaging electron density of cylindrical ion track with a radius of 0.36 {\AA} step by step. The trend of the axial force is similar to that of electron density, and there is a proportional relation between the two to some extent, which is in accordance with the density functional theory (DFT) results \cite{Echenique1986Nonlinear} for FEG. The average density values obtained along the whole trajectories in $<$100$>$, $<$110$>$, and $<$111$>$ channels are 0.35, 0.33, and 0.28 electrons/{\AA}$^{3}$, respectively, which consists with the amplitude of \emph{$S_{e}$} shown in Fig.~1(b). This suggests that the \emph{$S_{e}$} in channeling conditions is related to the average density along the projectile's trajectory, corroborating the interpretation in the literatures \cite{Wang1997Nonlinear,Wang1998Constrained,Winter2003Energy,Martingondre2013Scattering,PhysRevB.91.125203}.

The trend of force also have a direct correlation with the threshold effect. As can be seen, the applied force on helium ion
varies like a sine function with respect to the displacement, and trajectories with the low impact parameters have a larger variation. For trajectories 3 and 4, the force is generally above zero, i.e. positive $z$ direction, which means the ions are predominantly exposed to drag force along the channel. For trajectory 5, the force exerting on the ion
turns up and down periodically (in the $z$ axis direction), leading
to little net energy loss in the following path, which is consistent with the relatively small threshold velocity of trajectory 5 shown in Fig.~5(b).

It should be noted that, since projectile have distinct charge transfer behaviors along different trajectories, and charge transfer contribute significantly to \emph{$S_{e}$}. So the effect of impact parameter on \emph{$S_{e}$} is also embodied through the charge transfer.

\begin{figure}[htp]
  \centering
  \includegraphics[width=9cm,height=8cm]{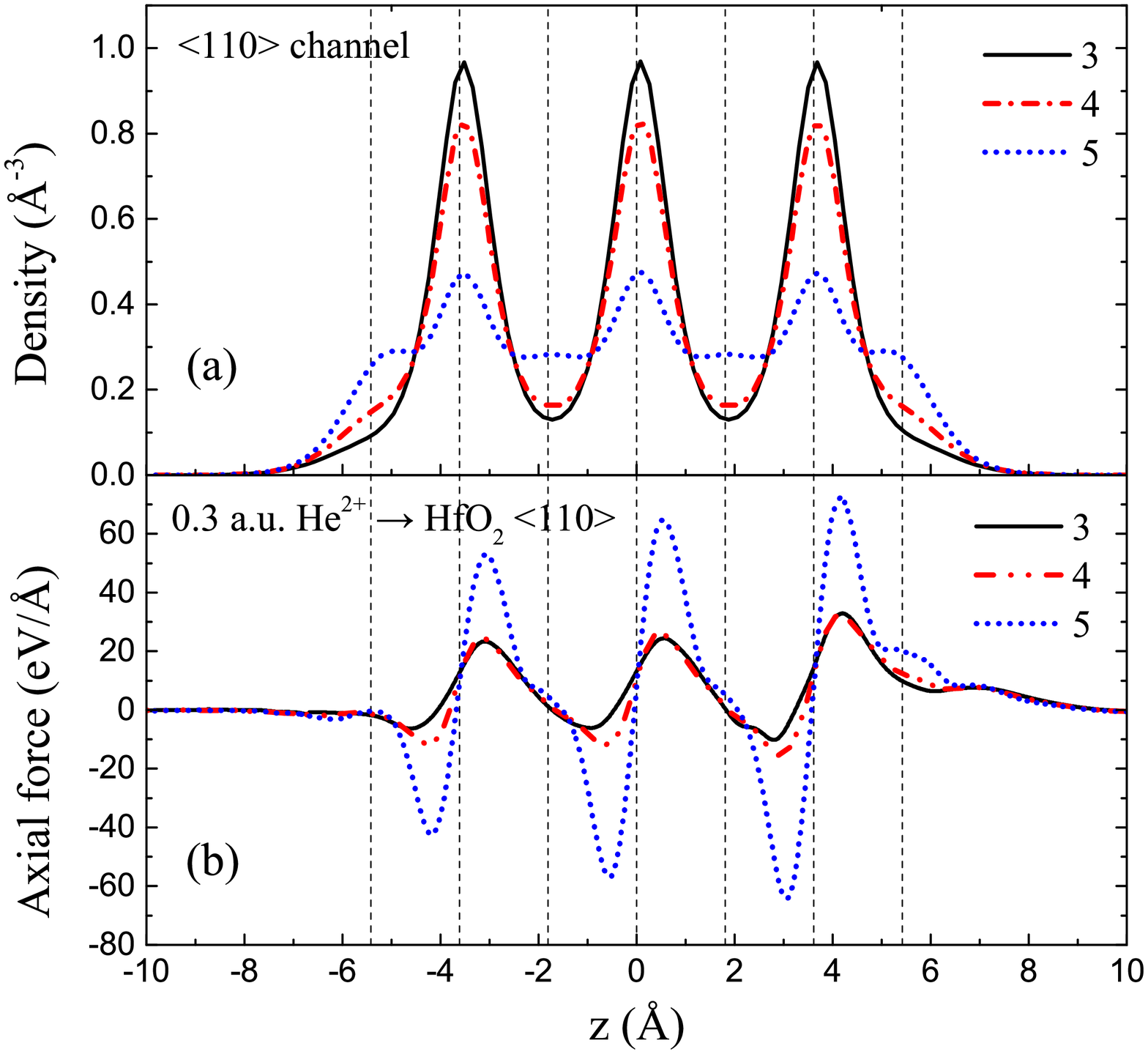}\\
  \caption{(Color online) Electron densities (a) and axial force (b) exerted on helium ion for a given velocity of 0.3 a.u. versus \emph{z} coordinates in HfO$_{2}$ along three trajectories in $<$110$>$. The curves are corresponding to trajectories in Fig.\,5(b). The vertical dashed lines are positions of crystalline layers comprising both O and Hf atoms.}
\end{figure}

\section{ CONCLUSIONS}

Theoretical study from first principles the electronic stopping power of slow light projectiles in HfO$_{2}$ has been presented.
The velocity--proportional \emph{$S_{e}$} of HfO$_{2}$ is predicted. A quantitative agreement between the experimental data and our results is achieved. Threshold effect is found when proton is channeled in the HfO$_{2}$ thin film, while, the expected threshold effect was not found for helium ion channelling the crystal, which was interpreted as a consequence of charge transfer. We have learned that the \emph{$S_{e}$} is sensitive to the impact parameter due to the different electron density experienced by the ions, which is consistent with the assumptions form FEG model. Our results shed light on describing the interaction between the ions and the target electrons without restricting the electrons to the adiabatic surface. To obtain a deeper understanding of the effect of charge transfer on inelastic energy loss, a thorough theoretical analysis of possible charge exchange mechanism in combination with suitable experimental studies is highly desirable.

\vskip 10mm

\begin{center}
\textbf{ACKNOWLEDGEMENTS}
\end{center}

One author wants to thank Dr.~F. Mao for the fruitful discussions. This work was supported by the National Natural Science Foundation of China under Grant Nos.\,11635003, 11025524 and 11161130520, National Basic Research Program of China under Grant No.\,2010CB832903, and the European Commissions 7th Framework Programme (FP7-PEOPLE-2010-IRSES) under Grant Agreement Project No.\,269131.

%

\end{document}